\documentclass[letterpaper, 10 pt, conference]{ieeeconf}  

\IEEEoverridecommandlockouts                              

\usepackage{amsmath, amsfonts, amssymb}
\usepackage{booktabs}
\usepackage{graphicx}
\usepackage{hyperref}
\usepackage{xcolor}
\usepackage{algorithm}
\usepackage{algorithmic}
\usepackage{float}
\usepackage{tikz}
\usepackage{tabularx}
\usepackage{array,booktabs}
\usepackage{cite}
\usepackage{pgfplots}
\pgfplotsset{compat=1.18}
\usepgfplotslibrary{groupplots}

\usepackage[deletedmarkup=sout,authormarkup=superscript]{changes}
\definechangesauthor[name={Lauren}, color=blue]{LB}
\definechangesauthor[name={Dominic}, color=red]{DLM}

\newcommand{\nseeds}{7}

\newcommand{\jmobest}{\textcolor{red}{[$J^*_{\text{MO,HF}}$]}}
\newcommand{\jmolf}{\textcolor{red}{[$J^*_{\text{MO,LF}}$]}}

\newcommand{\ilccv}{\textcolor{red}{[CV\%]}}
\newcommand{\mocv}{\textcolor{red}{[CV\%]}}

\renewcommand{\ilccv}{5.6}

\renewcommand{\jmolf}{57{,}676}     
\renewcommand{\jmobest}{127{,}329}
\renewcommand{\mocv}{6.4}

\usepackage{enumitem}
\usetikzlibrary{positioning, arrows.meta, shapes.geometric, fit, backgrounds}
\usetikzlibrary{calc}

\title{\LARGE \bf
Iterative Learning Control of the Cooling Rate in a Dual-Laser Powder Bed Fusion Process
}

\author{Lauren Bogo$^{1}$, Adam Clare$^{1}$, and Dominic Liao-McPherson$^{1}$%
\thanks{© 2026 IEEE.  Personal use of this material is permitted.  Permission from IEEE must be obtained for all other uses, in any current or future media, including reprinting/republishing this material for advertising or promotional purposes, creating new collective works, for resale or redistribution to servers or lists, or reuse of any copyrighted component of this work in other works.}
\thanks{The authors are with the University of British Columbia Department of Mechanical Engineering, 2054-6250 Applied Science Ln, Vancouver, BC V6T 1Z4, Canada. Email: \texttt{\{lauren.bogo, adam.clare, dominic.liao-mcpherson\}@ubc.ca} This work was supported by the Natural Sciences and Engineering Research Council of Canada (Reference \#\,RGPIN-2023-03257).}}%

\begin{document}

\maketitle
\thispagestyle{empty} 
\pagestyle{empty}
\begin{abstract}
The thermal history of the melt pool in laser powder bed fusion (LPBF) additive manufacturing processes governs the solidification microstructure and the mechanical properties of the resulting 3D-printed parts. Dual-laser systems offer additional degrees of freedom to control the cooling profile by reheating material behind the melt pool, but calibrating process parameters is challenging due to the complex physics of the process. We present an optimization-based iterative learning controller that determines optimal power, velocity, and offset settings by judiciously combining simulations and experiments: the model supplies search directions while feedback obtained from experiments on the real plant corrects for parameter errors, enabling convergence despite model inaccuracies. The approach is validated in simulation using a 
high-fidelity thermal model as a plant surrogate, with deliberate mismatches in absorption coefficient, latent heat treatment, and powder-bed effective conductivity between plant and model. Results show that the controller drives the plant cost down by over an order of magnitude and reaches a tight band of low-cost solutions across seeds, while model-only feedforward optimization stalls at a substantially higher plant cost despite appearing to converge on the surrogate.
\end{abstract}

\section{Introduction}
\label{sec:intro}

Metal additive manufacturing has emerged as a transformative technology for producing parts with complex geometries. E.g., in the aerospace and biomedical fields it has enabled the fabrication of complex, lightweight components with unprecedented design freedom~\cite{debroy2018additive}. However, parts produced by laser powder bed fusion (LPBF) typically exhibit poor as-built microstructure including columnar grains, anisotropic mechanical properties, and residual stress, requiring costly post-processing such as heat treatment before they can meet performance specifications~\cite{debroy2018additive}. These additional steps negate many of the efficiency gains offered by additive manufacturing. Controlling microstructure \emph{during} the build, rather than correcting it afterward, would improve efficiency and reduce cost.

\begin{figure}[ht]
    \centering
    \includegraphics[width=0.9\columnwidth]{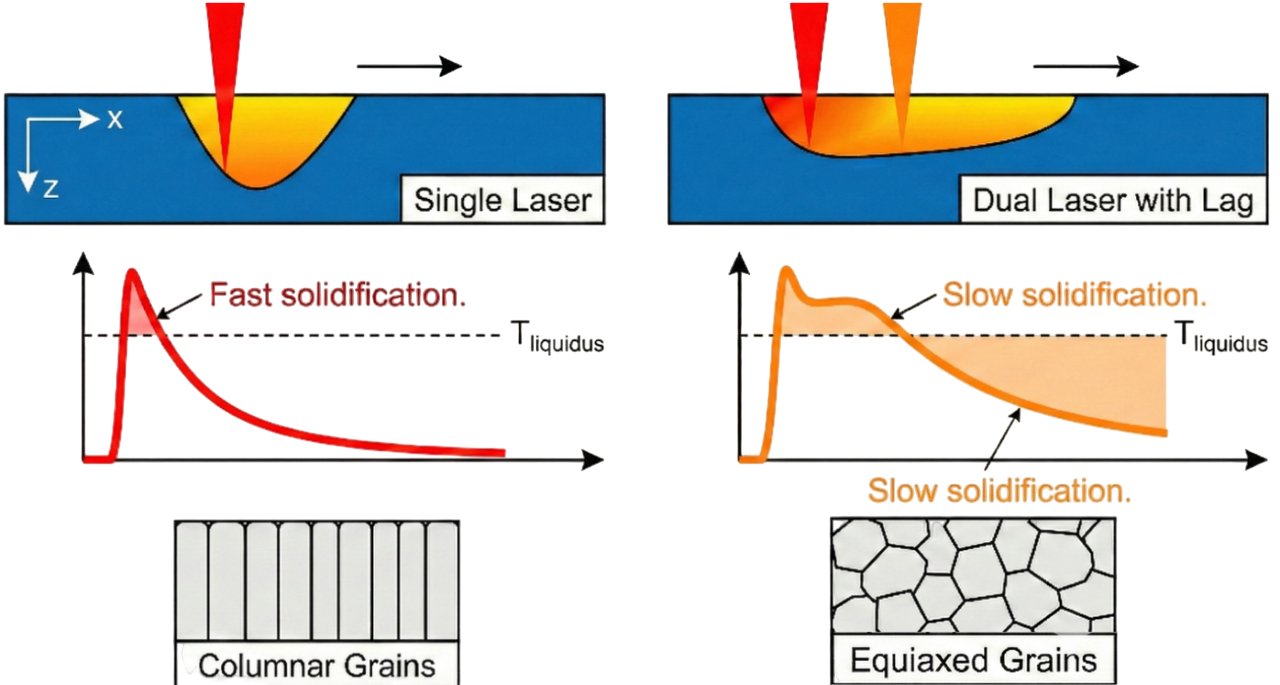}
    \caption{Dual-laser control of solidification microstructure. (Left) Single-laser processing produces steep thermal gradients and fast cooling, resulting in columnar grain growth. (Right) A trailing reheat laser reduces the cooling rate, extending time above the liquidus temperature and enabling the columnar-to-equiaxed transition. The resulting equiaxed microstructure exhibits improved isotropy and crack resistance.}
    \label{fig:microstructure}
\end{figure}

The mechanical properties of LPBF parts are governed by the thermal history experienced during solidification~\cite{chia2022process}. Temperature gradients on the order of $10^6$~K/m and cooling rates exceeding $10^6$~K/s drive microstructure formation, with a combination of the thermal gradient and solidification velocity determining whether columnar or equiaxed grain morphologies emerge~\cite{kurz2001columnar, hunt1984steady}. As illustrated in Figure~\ref{fig:microstructure}, steep gradients produce columnar dendritic structures with anisotropic properties and hot-cracking susceptibility, while reduced gradients enable equiaxed grains with isotropic, often superior mechanical performance~\cite{gaumann2001single}. 

The majority of LPBF machines use only a single laser which offers limited ability to modulate thermal history: once power and scan speed are set, the temperature field is governed by heat conduction into the substrate. Dual-laser configurations introduce additional degrees of freedom which can be used to decouple melting from thermal management. In dual-laser LPBF, a trailing secondary laser is used to reheat material during solidification to reduce cooling rates and thermal gradients~\cite{zhang2022process,bergmueller2023enhancing}, promoting the columnar-to-equiaxed transition (CET)~\cite{bergmueller2023enhancing,vanini2025local}. However, the additional degrees of freedom must be used correctly to reliably realize these benefits: incorrect parameters provide no improvement or can degrade part quality. To date, all reported dual-laser LPBF studies set trailing laser parameters using feedforward approaches, either through experimental parameter sweeps~\cite{bergmueller2023enhancing} or by selecting parameters based on simulations~\cite{zhang2022process, vanini2025local}. These procedures must be repeated for every new material, machine, or geometry, and the resulting parameters are only as good as the model or the experimental budget allows. Moreover, derivative-free alternatives scale unfavorably with input
dimension. A grid search over even a modest 10-point discretization
of the four-dimensional input space $(P_{L1}, P_{L2}, d, v)$ would
require $10^4$ physical experiments, and Bayesian optimization,
while more sample-efficient than grid search, suffers from the curse
of dimensionality and typically requires a substantial evaluation
budget at this input dimension~\cite{shahriari2015taking}. By
contrast, the approach presented here drives the plant cost down by
over an order of magnitude within roughly 20-40 plant evaluations by
exploiting model gradients to determine search directions while
using plant feedback to correct for model error.

In this paper, we present an algorithm for determining optimal process parameters for dual-laser LPBF process using optimization-based iterative learning control (OB-ILC). We optimize the leading and trailing laser powers, lag distance, and scan velocity to track target thermal profiles through the solidification region, promoting equiaxed grain formation. The controller uses a low-fidelity differentiable simulation to compute search directions via a data-driven sequential quadratic programming (SQP) algorithm, while experiments on the plant correct for model mismatch at each iteration. We validate the framework using JAX-FEM thermal simulations~\cite{xue2023jax} where the plant and model differ in absorption coefficient, latent heat treatment, and powder-bed effective conductivity, representing the realistic scenario where a fast surrogate omits physics present in the true process. Results show that ILC converges to consistent low-cost plant solutions across initial conditions, while model-only optimization, despite appearing to converge on the surrogate, settles at significantly higher plant cost. The gap between the two reflects the model-plant mismatch that the iterative feedback loop corrects. Because the plant is treated as a black box, the same algorithm applies directly to hardware experiments: replacing the high-fidelity simulation with pyrometer measurements requires no algorithmic changes, which allows for systematic calibration on real machines without extensive manual tuning.

ILC is well established in the manufacturing domain~\cite{bristow2006survey,wang2009survey} where it has a natural synergy with the inherently repetitive nature of many manufacturing processes. ILC was originally formulated as a discrete-time tracking problem in the iteration domain; OB-ILC extends the paradigm by formulating the ILC problem as a constrained optimization problem~\cite{owens2005iterative} which provides a natural framework for integrating constraints and more complex objectives and for obtaining convergence/robustness guarantees~\cite{mishra2010optimization, liao2022robustness}. Prior applications of ILC to AM include passivity-based temperature tracking in selective laser melting~\cite{spector2018passivity}, model-free multi-objective thermal regulation~\cite{inyang2022model}, melt pool geometry control~\cite{liao2024layer}, layer-to-layer thermal regulation~\cite{kavas2025layer}, and spatial height control~\cite{balta2024iterative}. To the best of the authors' knowledge, this work represents the first application of ILC to dual-laser thermal profile optimization in laser powder bed fusion.

The remainder of this paper is organized as follows. Section~\ref{sec:problem} formulates the problem. Section~\ref{sec:ilc} presents the multi-fidelity ILC framework. Section~\ref{sec:simulation} describes the simulation study. Section~\ref{sec:results} presents single-line geometry results. Section~\ref{sec:results:zigzag} tests transfer to a non-single-line (zigzag) geometry. Section~\ref{sec:conclusion} concludes.

\section{Problem Formulation}
\label{sec:problem}

We consider a dual-laser powder bed fusion system where two lasers traverse a shared straight-line scan path at velocity $v$. A leading laser creates the melt pool while a trailing laser reheats material during solidification to modulate the cooling rate. We study a single straight-line geometry as the fundamental building block: most LPBF scan strategies employ hatching patterns composed of straight lines, so calibrating the trailing laser for a single line is a natural starting point. We focus on layer-to-layer control, where trailing laser parameters are updated between trials rather than within a scan. Within-scan control would require high-bandwidth thermal monitoring hardware; layer-to-layer control runs on existing machines.

The thermal response of the powder bed is governed by the transient heat equation on a three-dimensional domain $\Omega \subset \mathbb{R}^3$:
\begin{equation}
    \rho \, c_p(T) \frac{\partial T}{\partial t} = \nabla \cdot (k \nabla T) + q(x, t),
    \label{eq:heat}
\end{equation}
where $\rho$ is density, $c_p(T)$ is the (temperature-dependent) specific heat, $k$ is thermal conductivity, and $q(x,t)$ is the volumetric heat source. The boundary conditions are
\begin{align}
    T \big|_{z=0} &= T_0, \label{eq:bc_bottom} \\
    -k \frac{\partial T}{\partial n}\bigg|_{\text{top}} &= q_{\text{laser}} - h(T - T_\infty) - \sigma \varepsilon (T^4 - T_\infty^4), \label{eq:bc_top} \\
    -k \frac{\partial T}{\partial n}\bigg|_{\text{walls}} &= -h(T - T_\infty) - \sigma \varepsilon (T^4 - T_\infty^4), \label{eq:bc_walls}
\end{align}
where $T_0 = T_\infty = 300$~K, $h = 100$~W/(m$^2\cdot$K) is the convection coefficient, and $\varepsilon = 0.3$ is the surface emissivity. The bottom face acts as the primary heat sink, while the top and side walls lose heat by convection and radiation.

The heat source is the superposition of two Gaussian laser beams where each beam deposits energy as
\begin{equation}
    q_i(x, t) = \frac{\eta P_i}{\pi r_b^2 \, \delta_z} \exp\!\Bigl(-\frac{\|x_{xy} - \mu_i(t)\|^2}{r_b^2}\Bigr).
    \label{eq:gaussian}
\end{equation}
Here $\eta$ is the laser absorption coefficient, $P_i$ is the power of laser $i$, $r_b$ is the beam radius, $\delta_z$ is the optical penetration depth, $x_{xy}$ denotes the in-plane coordinates, and $\mu_i(t)$ is the beam centre position at time $t$. The leading laser moves along the scan direction $\hat{e}$ at velocity $v$ with $\mu_1(t) = v t \, \hat{e}$, and the trailing laser follows at a fixed lag distance $d$ behind: $\mu_2(t) = \mu_1(t) - d \, \hat{e}$.
The control input is
\begin{equation}
    u = \begin{bmatrix} P_{L1} \\ P_{L2} \\ d \\ v 
    \end{bmatrix} \in \mathcal{U} \subset \mathbb{R}^4,
    \label{eq:input}
\end{equation}
where $P_{L1} \in [25, 100]$~W and $P_{L2} \in [25, 100]$~W 
are the leading and trailing laser powers, $d \in [0, 250]~\mu$m 
is the lag distance, and $v \in [0.25, 1.0]$~m/s is the scan 
velocity. All four parameters influence the thermal history: the laser 
powers govern energy input, the lag distance controls the timing 
of the reheat pulse relative to solidification, and the scan 
velocity sets the interaction time between the lasers and the 
powder bed.

\begin{figure}[ht]
    \centering
    \includegraphics[width=0.5\columnwidth]{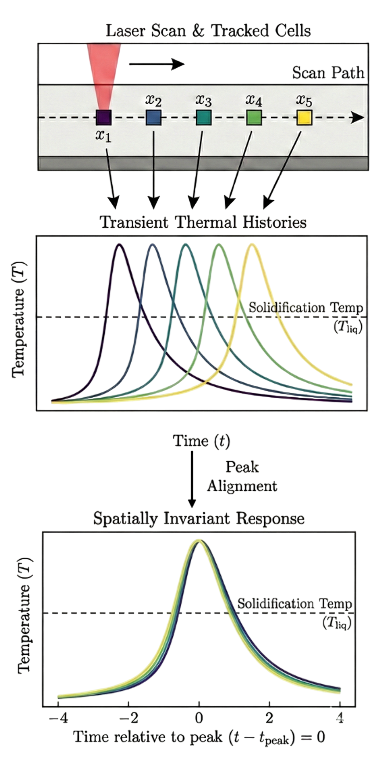}
    \caption{Spatial invariance: cells at different positions along the scan path experience identical thermal responses shifted in time. Peak-aligned responses collapse to a single curve.}
    \label{fig:invariance}
\end{figure}

The system output is the temperature at $m$ control points 
sampled during the cooling phase at a single cell on the scan 
path centreline:
\begin{equation}
    y = \begin{bmatrix} T(\tau_1) & T(\tau_2) & \cdots & 
    T(\tau_m) \end{bmatrix}^T \in \mathbb{R}^{m},
\end{equation}
where $\tau_j = t_0 + \Delta t_j$ and $t_0$ is the last time 
$T(t)$ crosses $T_\text{liq}$ from above, marking the onset of 
final solidification. This anchoring ensures the control points 
capture the cooling history that determines microstructure, 
regardless of any trailing-laser-induced remelting. The tracked cell is located in the middle of the scan path. Spatial invariance along the scan path (Figure~\ref{fig:invariance}) ensures that outside of a small boundary region cells at different positions experience essentially identical thermal responses shifted in time, so optimizing at a single interior cell is effectively identical to optimizing the entire track.


For a given input $u$ we can obtain a thermal response $y = P(u)$ from the plant $P: \mathbb{R}^n \to \mathbb{R}^{m}$ by \emph{running an experiment.} To help us optimize the process, we also have a simulation $M: \mathbb{R}^n \to \mathbb{R}^{m}$ that approximates $P$ by solving \eqref{eq:heat} numerically and can provide us with the Jacobian $J = \nabla M \in \mathbb{R}^{m \times n}$ via reverse-mode automatic differentiation in JAX-FEM~\cite{xue2023jax}.

\begin{figure}[ht]
    \centering
    \includegraphics[width=0.7\columnwidth]{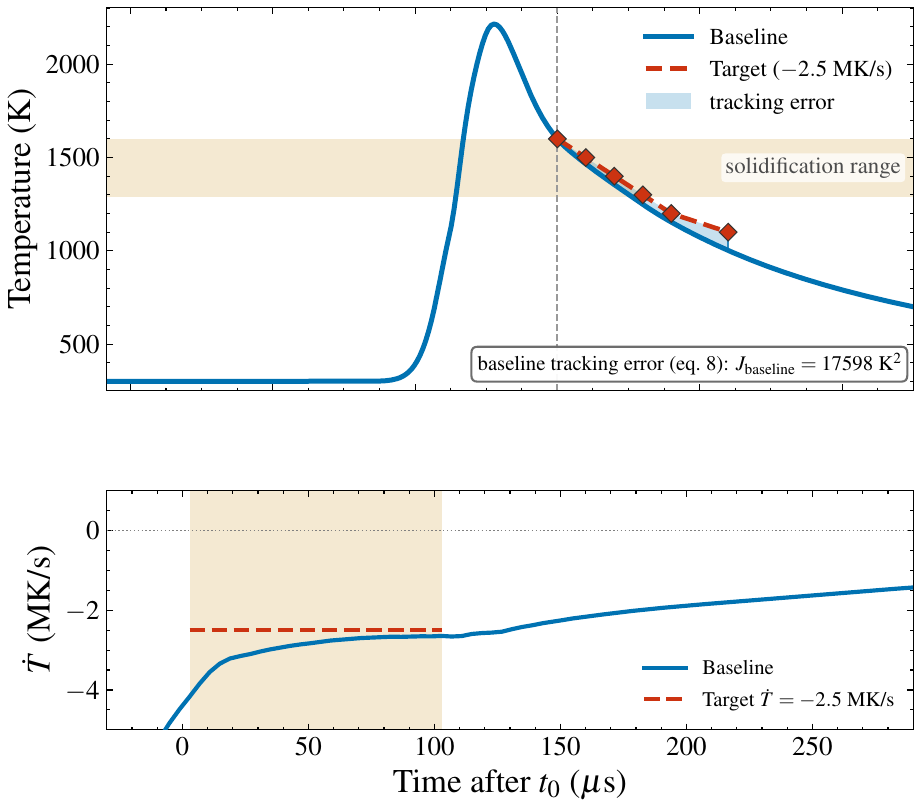}
    \caption{Control point tracking objective. The single-laser baseline
thermal response ($P_{L1}=50$~W, $P_{L2}=0$) evaluated on the plant
heats above the liquidus, peaks near 2950~K, and cools rapidly back
through the solidification region (shaded). Target temperatures $T_j^*$ (red
diamonds) prescribe a cooling profile favouring equiaxed grains at
$\dot{T}_{\text{target}} = -2.5$~MK/s from $T_\text{liq}$ at $t_0$.}
\label{fig:objective}
\end{figure}

The target is a cooling profile through the solidification range, from
the solidus $T_\text{sol}=1290$~K to the liquidus $T_\text{liq}=1600$~K
(shaded in Figure~\ref{fig:objective}). Fast cooling, typical of
single-laser LPBF, produces columnar grains, whereas slower cooling
favours equiaxed grains~\cite{kurz2001columnar, hunt1984steady, gaumann2001single}; the
trailing laser slows cooling by reheating behind the melt
pool~\cite{bergmueller2023enhancing}. LPBF solidification rates are on
the order of $10^6$~K/s~\cite{kurz2001columnar, hunt1984steady,marchese2024heat}, so we set a target cooling rate
$\dot{T}_\text{target} = -2.5$~MK/s, slower than a single-laser scan, as
a representative microstructure-relevant setpoint rather than a precise
transition threshold. The profile is encoded by $m = 6$ control points
at offsets $\Delta t_j \in \{0, 40, 80, 120, 160, 240\}~\mu$s after
$t_0$, with targets $T_j^* = \{1600, 1500, 1400, 1300, 1200, 1100\}$~K,
four of which lie within the solidification range.

\section{Iterative Learning Controller}
\label{sec:ilc}
Our control objective is to determine $u \in \mathcal{U}$ that minimizes the tracking error
\begin{equation}
    f(u) = \sum_{j=1}^{m} \bigl(T(\tau_j;\, u) - T_j^*\bigr)^2,
    \label{eq:objective}
\end{equation}
where $T(\tau_j;\, u)$ are the temperature components of $y = P(u)$ and $T_j^*$ are the target temperatures at the $m$ control points, as shown in Figure~\ref{fig:objective}.

Optimization-based ILC solves $\min_{u \in \mathcal{U}} f(u)$ iteratively by combining model gradients with plant feedback. We introduce the predicted output $\hat{y} \in \mathbb{R}^{m}$ as a decision variable and recast the problem as
\begin{subequations} \label{eq:opt_problem}
\begin{align}
    \min_{u, \hat{y}} \quad & f(\hat{y}) \label{eq:opt_obj} \\
    \text{subject to} \quad & \hat{y} = P(u), \label{eq:opt_constraint}
\end{align}
\end{subequations}
where $f(\hat{y}) = \sum_j (\hat{y}_j - T_j^*)^2$. Following~\cite{balula2023sequential}, we apply an iteration of sequential quadratic programming (SQP) to~\eqref{eq:opt_problem}, replacing the unavailable Jacobian of the true plant $\nabla P$ with the model Jacobian $J = \nabla M$. The resulting QP subproblem is:
\begin{subequations} \label{eq:qp}
\begin{align}
    \min_{\Delta u, \Delta \hat{y}} \quad &
    \frac{1}{2} \begin{pmatrix} \Delta u \\ \Delta \hat{y} \end{pmatrix}^{\!T} B_k
    \begin{pmatrix} \Delta u \\ \Delta \hat{y} \end{pmatrix} + \nabla f(\hat{y}_k)^T \Delta \hat{y} \label{eq:qp_obj} \\
    \text{s.t.} \quad & \Delta \hat{y} - J_k \, \Delta u = y_k - \hat{y}_k \label{eq:qp_constraint}\\
    & u_k + \Delta u \in \mathcal{U}
\end{align}
\end{subequations}
where $B_k$ is a Hessian approximation, $J_k = \nabla M(u_k) \in \mathbb{R}^{m \times n}$ is the model Jacobian, and $\nabla f$ is the gradient of the cost with respect to the predicted output. Each iteration requires one plant evaluation and one model Jacobian computation. The constraint~\eqref{eq:qp_constraint} is a linearization of $\hat{y} = P(u)$ around $(u_k, \hat{y}_k)$, with the right-hand side $y_k - \hat{y}_k$ correcting for model mismatch at each iteration.

Substituting the linearized constraint~\eqref{eq:qp_constraint} into the
objective~\eqref{eq:qp_obj} eliminates $\Delta\hat y$ and yields a
quadratic program in the $n$ design variables alone. With
$\Delta\hat y = J_k\,\Delta u + (y_k - \hat y_k)$, the predicted output
after the step is $\hat y_k + \Delta\hat y = y_k + J_k\,\Delta u$, so
the objective reduces to $\lVert J_k\,\Delta u + e_k\rVert^2$ with
$e_k = y_k - T^* \in \mathbb{R}^m$ the tracking error. The resulting
\emph{reduced QP} is
\begin{equation}
\begin{aligned}
    \min_{\Delta u}\quad & \tfrac{1}{2}\,\Delta u^\top B_{x,k}\,\Delta u
    + g_{x,k}^\top \Delta u \\
    \text{s.t.}\quad & u_k + \Delta u \in \mathcal{U},
\end{aligned}
\label{eq:reduced_qp}
\end{equation}
where
\begin{equation}
    g_{x,k} = J_k^\top e_k \label{eq:reduced_grad}
\end{equation}
is the reduced gradient and $B_{x,k} \in \mathbb{R}^{n\times n}$
approximates the reduced Hessian. Formulation~\eqref{eq:reduced_qp} is
equivalent to~\eqref{eq:qp} but is cheaper to solve and requires only an
$n\times n$ BFGS update; we maintain $B_{x,k}$ directly and never form
the full-space Hessian $B_k$. It also exposes the multi-fidelity structure
directly: the search direction is built from the model Jacobian $J_k$,
while the residual $e_k$ is set by the plant output $y_k$, so
plant feedback enters the gradient at every iteration.

The reduced QP~\eqref{eq:reduced_qp} is solved for the step $\Delta u$.

We refine the reduced Hessian $B_{x,k}$ each iteration with a BFGS update built
from the design step $s_k = u_{k+1} - u_k$ and the gradient change
$p_k = g_{x,k+1} - g_{x,k}$:
\begin{equation}
    B_{x,k+1} = B_{x,k} + \frac{p_k p_k^\top}{p_k^\top s_k}
    - \frac{B_{x,k}\, s_k s_k^\top B_{x,k}}{s_k^\top B_{x,k}\, s_k}.
    \label{eq:rbfgs}
\end{equation}
When the curvature $s_k^\top p_k$ falls below $0.2\, s_k^\top B_{x,k} s_k$, we
apply Powell's damping: $p_k$ is replaced by
$\tilde p_k = \theta p_k + (1-\theta) B_{x,k} s_k$, with $\theta \in (0,1)$
chosen so that $s_k^\top \tilde p_k = 0.2\, s_k^\top B_{x,k} s_k$, which keeps
$B_{x,k+1}$ positive definite.

\begin{figure}[ht]
    \centering
    \includegraphics[width=0.75\columnwidth]{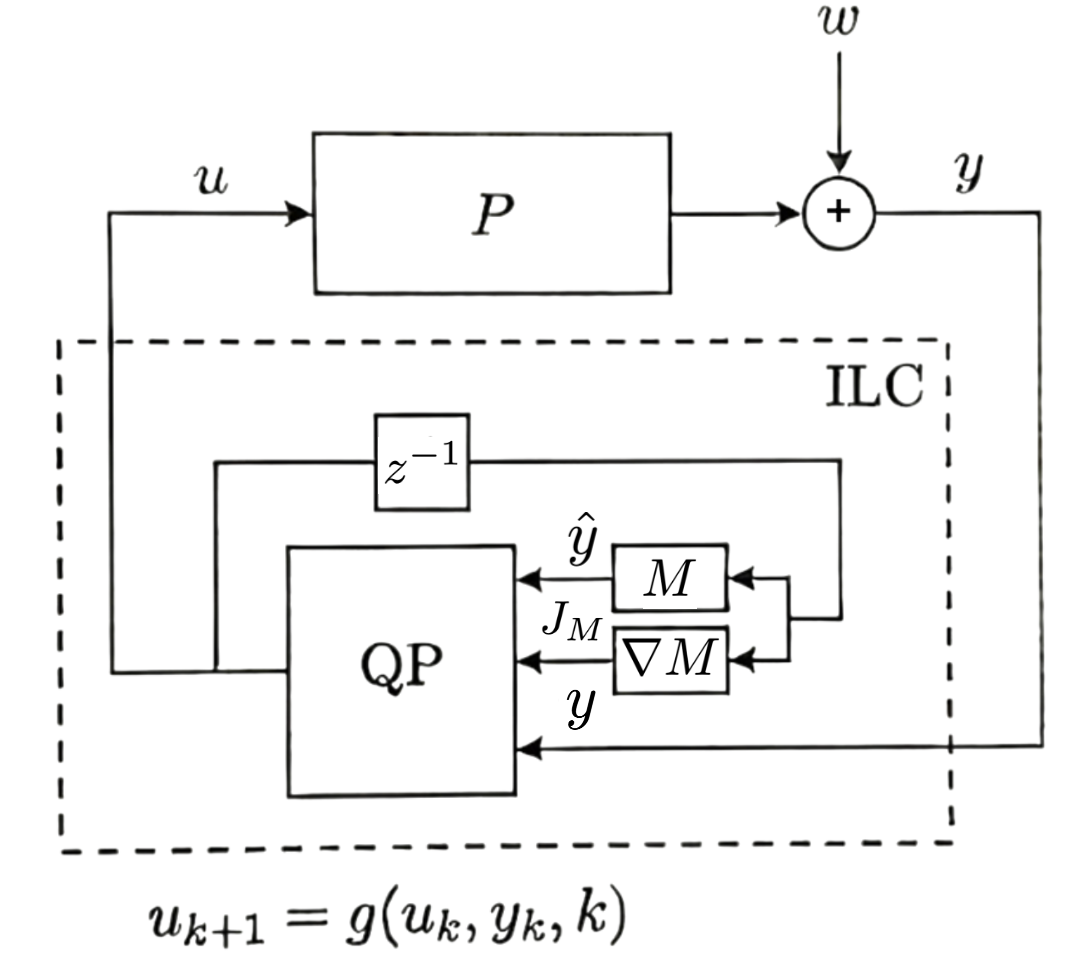}
    \caption{ILC as a closed-loop controller in the iteration domain. The plant (top) is evaluated once per iteration. The compensator (dashed) uses $M$ and $J = \nabla M$ to solve a QP, yielding $u_{k+1} = g(u_k, y_k, k)$. The model determines where to step; the plant measurement determines how far off we are. The block $z^{-1}$ denotes the unit delay in the iteration domain: $z^{-1} u_{k+1} = u_k$.}
    \label{fig:block_diagram}
\end{figure}

This procedure yields a compensator
\begin{equation}
    u_{k+1} = g(u_k, y_k, k),
    \label{eq:compensator}
\end{equation}
which is placed in closed-loop with the plant
\begin{equation}
    y_k = P(u_k) + w_k
\end{equation}
where $w_k$ represents process/measurement noise, as illustrated in Figure~\ref{fig:block_diagram}. Because additive manufacturing is inherently repetitive, this trial-based formulation is natural: each iteration corresponds to a scan, and the controller updates parameters between trials.

The complete procedure is summarized in Algorithm~\ref{alg:mfilc}. Model-only optimization provides the warm start $u_0$; subsequent iterations refine using plant feedback.

\begin{algorithm}[h]
\caption{Multi-Fidelity SQP-Based ILC}
\label{alg:mfilc}
\begin{algorithmic}[1]
\REQUIRE Plant $P$, Model $M$, Cost $f$, target $T^*$, tolerance $\epsilon$
\STATE Initialize: solve $\min_u f(M(u))$ for $u_0$; set $B_{x,0} = I$
\REPEAT
    \STATE Evaluate plant: $y_k = P(u_k)$
    \STATE Compute model Jacobian: $J_k = \nabla M(u_k)$
    \STATE Compute reduced gradient: $g_{x,k} = J_k^\top (y_k - T^*)$
    \STATE Solve reduced QP~\eqref{eq:reduced_qp} for $\Delta u$
    \STATE Update: $u_{k+1} = u_k + \Delta u$
    \STATE Update $B_{x,k+1}$ via reduced-space BFGS~\eqref{eq:rbfgs}
\UNTIL{$|f(u_k) - f(u_{k-1})| < \epsilon$}
\RETURN $u_k$
\end{algorithmic}
\end{algorithm}

\section{Simulation Study}
\label{sec:simulation}

We validate the framework using two JAX-FEM thermal simulations~\cite{xue2023jax}.
The high-fidelity (HF) simulation serves as the plant $P$, providing
temperature evaluations but no gradients to the optimizer. The
low-fidelity (LF) simulation serves as the model $M$, providing the
Jacobian $J = \nabla M$ via reverse-mode automatic differentiation but
containing deliberate physical mismatch representative of real modelling
uncertainty. Both solve~\eqref{eq:heat} with Inconel~625 properties
(Table~\ref{tab:material}) and beam radius $r_b = 50~\mu\mathrm{m}$.
The algorithm is initialized from \nseeds{} random starting points
sampled across $\mathcal{U}$ to account for non-convexity and to
characterize convergence reliability.

\begin{table}[ht]
\centering
\caption{Material properties for Inconel~625.}
\label{tab:material}
\begin{tabular}{lcc}
\toprule
Property & Value & Units \\
\midrule
Thermal conductivity $k$ (solid) & 15.0 & W/(m$\cdot$K) \\
Density $\rho$ & 8440 & kg/m$^3$ \\
Specific heat $c_p$ & 588 & J/(kg$\cdot$K) \\
Liquidus temperature $T_{\text{liq}}$ & 1600 & K \\
Solidus temperature $T_{\text{sol}}$ & 1290 & K \\
Latent heat of fusion $L_f$ & 227 & kJ/kg \\
\bottomrule
\end{tabular}
\end{table}

Table~\ref{tab:model_config} summarizes the two configurations. We
introduce mismatch on three quantities that are genuinely uncertain in
practice for LPBF. The effective absorptivity governing laser energy coupling is uncertain in LPBF, varying with powder state and melt-pool regime~\cite{trapp2017situ}; we use $\eta$ = 0.25 (plant) and 0.20 (model) as representative values, introducing the difference as model mismatch. Latent heat of
fusion is included in the plant via the apparent-heat-capacity method,
which raises the specific heat by $L_f/(T_\text{liq}-T_\text{sol})$
across the freezing range (more than doubling it in the solidification
region), and is omitted from the model. The effective thermal conductivity of the powder bed is an uncertain
parameter in LPBF, depending strongly on bed porosity~\cite{zhang2019thermal}. The HF model uses $k = 8$~W/(m$\cdot$K) as a
porosity-corrected effective value, while the LF model uses the bulk
solid value $k = 15$~W/(m$\cdot$K), reflecting the common
fast-surrogate simplification of treating the build region as fully
consolidated. The HF model does not include a vaporization sink, so peak temperatures above approximately 3000~K should be read as numerical artifacts rather than physical predictions. The cost function and reported metrics are evaluated in the solidification region (1290 to 1600~K), which is governed by conduction and is unaffected by the missing vapour physics.

\begin{table}[ht]
\centering
\caption{Model configurations for plant (HF) and surrogate (LF).}
\label{tab:model_config}
\footnotesize
\setlength{\tabcolsep}{3pt}
\begin{tabularx}{\columnwidth}{@{}Xcc@{}}
\toprule
Parameter & HF (Plant) & LF (Model) \\
\midrule
Absorption coeff. $\eta$ & 0.25 & 0.20 \\
\addlinespace
Latent heat & Active ($c_{p,\text{eff}}$) & Off \\
\addlinespace
Powder-bed $k$ (W/(m$\cdot$K)) & 8.0 & 15.0 \\
\addlinespace
Mesh (relative) & Full & $2.0\times$ coarser \\
\bottomrule
\end{tabularx}
\end{table}

\begin{figure}[ht]
    \centering
    \includegraphics[width=0.9\columnwidth]{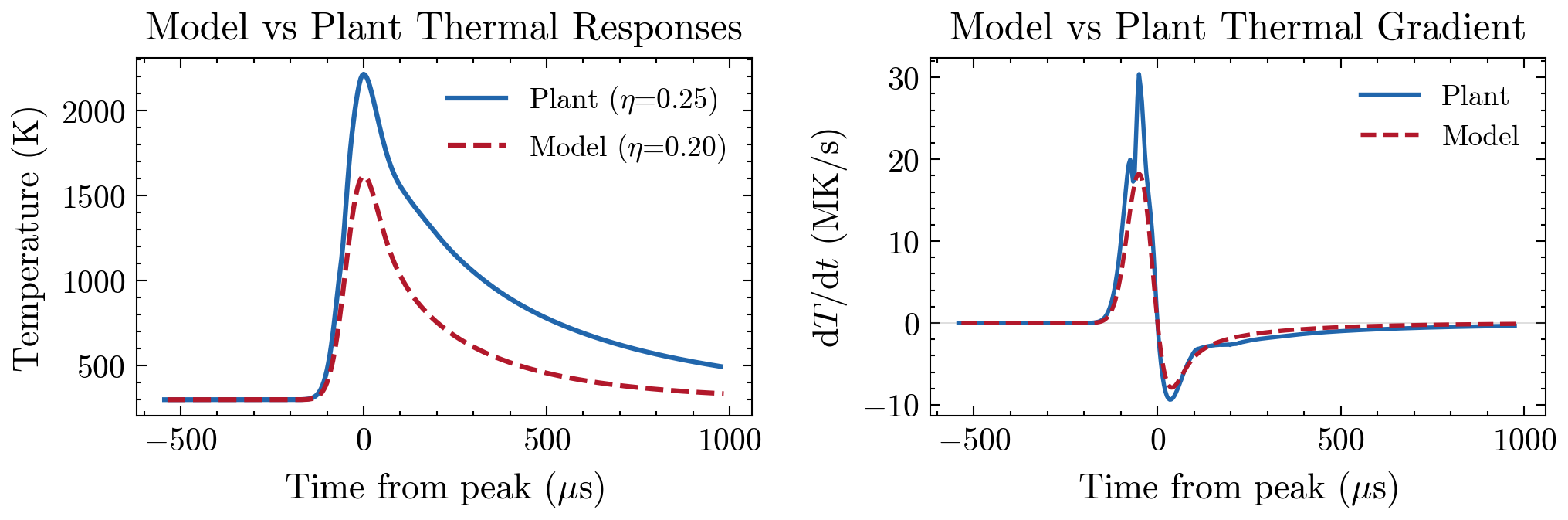}
    \caption{Thermal response (left) and cooling rate (right) of plant
    (HF: $\eta = 0.25$, $k=8$, latent heat on) and model
    (LF: $\eta = 0.20$, $k=15$, no latent heat) at identical input. The
    plant exhibits both reduced peak temperature and structurally
    different cooling rate through the solidification region (shaded).}
    \label{fig:mismatch}
\end{figure}

Figure~\ref{fig:mismatch} illustrates the cumulative effect of these
mismatches: for identical leading-laser-only input, the plant and the
model produce thermal responses that differ in both peak temperature
and cooling-rate structure through the solidification region. Parameters
optimized on the LF model alone are not guaranteed to be optimal, or
even good, when applied to the plant.

We compare two cases, both evaluated on the plant:
\begin{enumerate}[itemsep=2pt, topsep=4pt]
    \item \textbf{Model-only (MO)}: minimize $f(M(u))$ on the LF model,
    apply result to the plant without feedback.
    \item \textbf{ILC}: optimize using LF gradients with plant feedback
    (Algorithm~\ref{alg:mfilc}).
\end{enumerate}
The MO case represents current practice: parameters are tuned offline on a simulation and applied open loop. ILC instead corrects for model mismatch iteratively using plant measurements, mimicking operation on hardware with thermal monitoring.

\section{Single-line geometry: results}
\label{sec:results}

We evaluate the controller on the powder-bed mismatch configuration
described in Section~\ref{sec:simulation}, with \nseeds{} initial
conditions for each of ILC and MO.

\subsection{Initial conditions (seeds) and convergence}

\begin{figure}[ht]
    \centering
    \includegraphics[width=0.95\columnwidth]{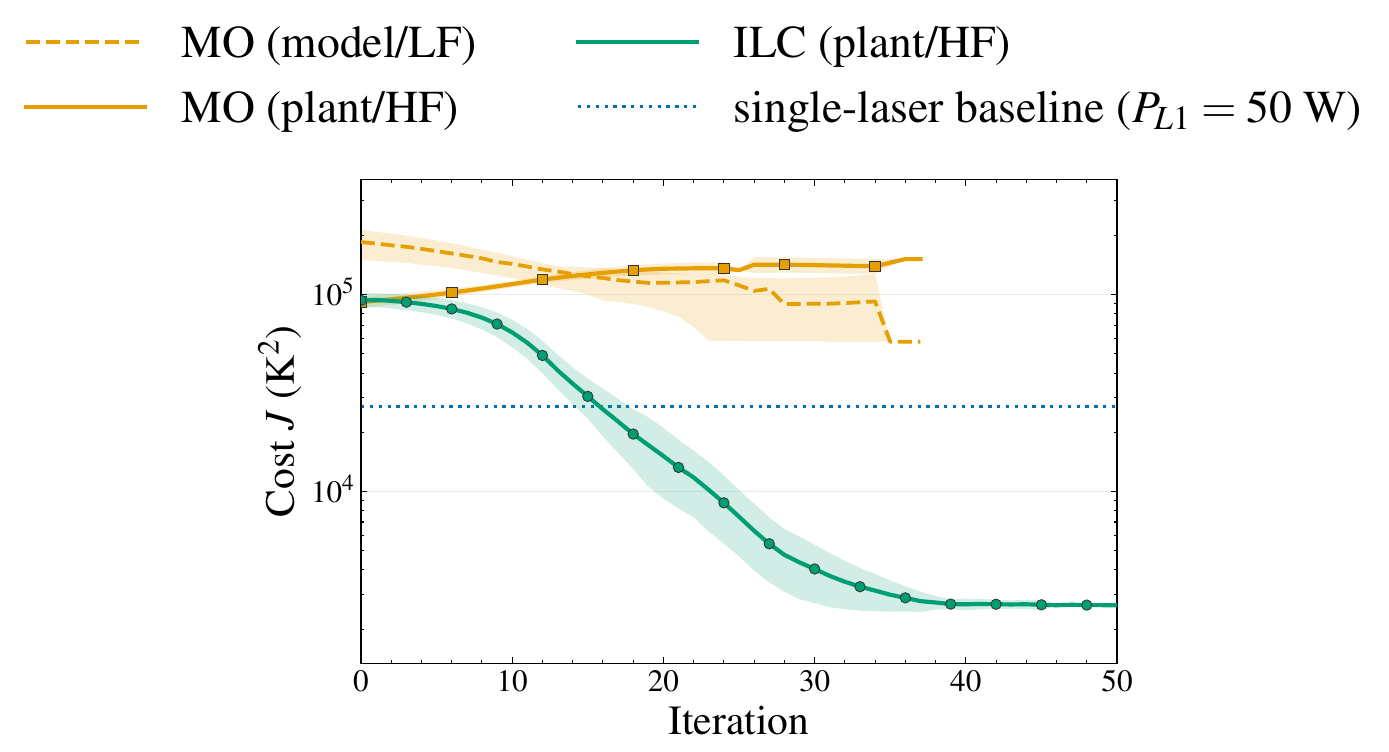}
    \caption{Convergence of cost $J(u_k)$ over \nseeds{} initial
conditions, shaded envelopes across seeds. ILC (green) uses plant
feedback at each iteration and decreases by an order of magnitude.
MO's LF cost (orange dashed) descends monotonically; its HF reeval
(orange solid) plateaus far above ILC. Blue dotted: single-laser
baseline at $P_{L1}=50$~W. The vertical gap between MO's LF and HF
traces is the model-plant mismatch ILC closes.}
    \label{fig:convergence}
\end{figure}

We refer to each random starting point $u_0 \in \mathcal{U}$ as a
\emph{seed}: a seed is just an initial condition for both ILC and MO,
sampled across $\mathcal{U}$ to characterize how the controller
behaves under different starting points. Figure~\ref{fig:convergence}
shows convergence of the plant cost $J(u_k)$ across iterations for
both methods. The ILC trajectory uses plant feedback at every step and
converges to a tight band across \nseeds{} seeds.
The MO trajectory differs between fidelities: the LF
cost (dashed) descends monotonically while MO optimizes against its
own model, but the HF cost (solid) at the same iterates stays well
above the ILC band. MO's best plant cost is \jmobest{} with
coefficient of variation \mocv{}, while its LF cost reaches \jmolf{}.
The gap between MO's LF and HF traces is the model-plant mismatch;
ILC closes it iteratively by using plant measurements to correct the
LF gradients.

\subsection{Convergence history at a single seed}
\label{sec:results:perseed}

\begin{figure}[ht]
    \centering
    \includegraphics[width=0.95\columnwidth]{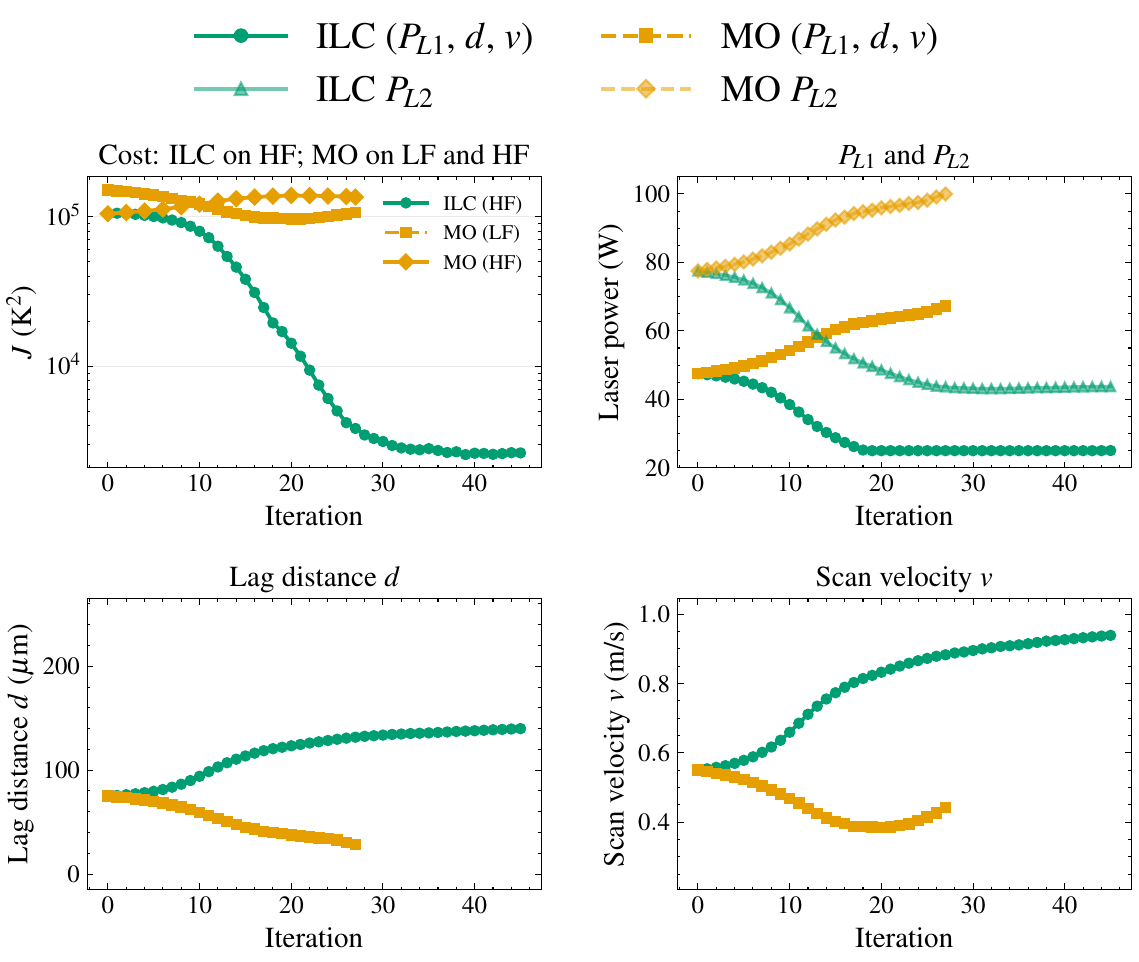}
    \caption{Per-seed dynamics (seed~7). ILC (green) and MO (orange)
    overlaid. Top-left: cost $J$ -- ILC on the HF plant (solid green),
    MO on the LF model (dashed orange), and MO re-evaluated on the HF
    plant at the iterates where that data is available (solid orange).
    Top-right: $P_{L1}$ and $P_{L2}$ (W). Bottom-left: lag $d$
    ($\mu$m). Bottom-right: scan velocity $v$ (m/s). ILC and MO
    converge to qualitatively different regions of the design space.}
    \label{fig:perseed}
\end{figure}

Figure~\ref{fig:perseed} shows the convergence at a single
representative seed (seed~7), the seed for which we currently have
the most complete MO HF reeval data. ILC reduces the plant cost by
over an order of magnitude; MO descends modestly on its LF objective
and plateaus, and its plant cost (HF reeval) stays at roughly
$10^5$~K$^2$ throughout. The two methods reach qualitatively
different parameter regions. ILC settles on a fast scan with the
leading-laser power at its lower bound, reaching the target by
removing energy. MO settles on a hot configuration with the trailing
laser near its upper bound and a slow scan, reaching its LF target
by adding energy. That strategy overshoots on the plant
(Section~\ref{sec:results:thermal}).

\subsection{Thermal profile tracking}
\label{sec:results:thermal}

\begin{figure}[ht]
    \centering
    \includegraphics[width=0.95\columnwidth]{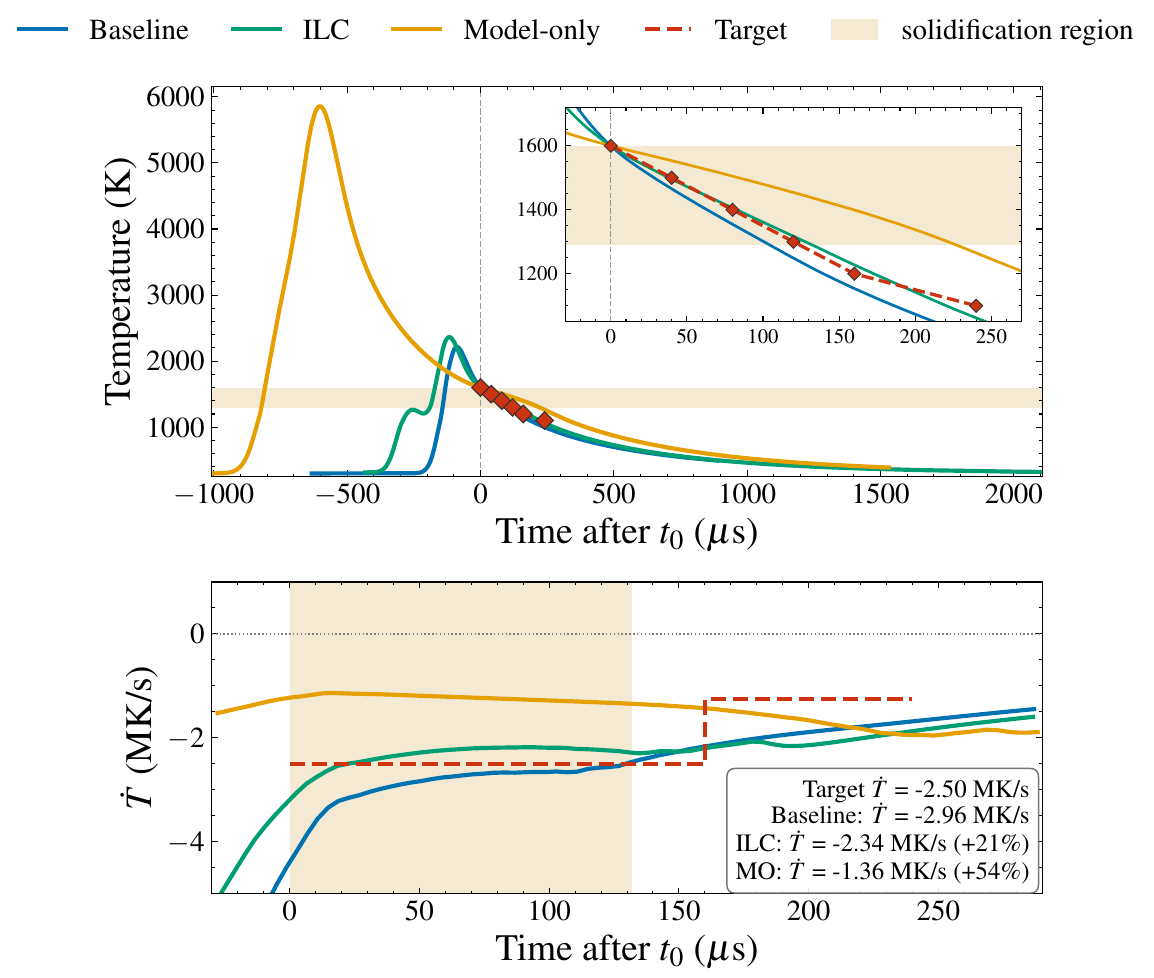}
    \caption{Thermal-profile tracking at the tracked cell on the
plant for seed~7. Top: $T(t)$ aligned at $t_0$ for baseline (blue,
$P_{L1}=50$~W, $P_{L2}=0$, $v=0.50$~m/s), ILC (green, $25.0$/$43.5$~W,
$d=138~\mu$m, $v=0.92$~m/s), and MO (orange, $63.8$/$96.3$~W,
$d=37~\mu$m, $v=0.39$~m/s); target control points (red diamonds,
dashed) prescribe $-2.5$~MK/s through the solidification region
(shaded). Bottom: cooling rate $\dot T(t)$ with target step-rate
(dashed red). ILC tracks the target; baseline cools too fast and
drops below; MO -- driven by LF model mismatch -- holds the cell
hot at roughly half the target rate.}
    \label{fig:profiles}
\end{figure}

Figure~\ref{fig:profiles} compares the thermal response at the
tracked cell, evaluated on the plant, for three cases at seed~7:
baseline single-laser ($P_{L2}=0$), ILC-optimized, and MO-optimized.
ILC tracks the target control points (red diamonds) closely through
the solidification region, with a mean cooling rate of $-2.34$~MK/s
against the target $-2.50$~MK/s (a deviation of 6\%). The
single-laser baseline cools too fast (mean $-2.96$~MK/s, 18\%
faster than target) and drops below the target profile shortly
after $t_0$.
The MO-optimized trace differs from the target qualitatively, cooling at $-1.36$~MK/s (roughly half the
target rate) and stays well above the target through the
solidification region.

MO and ILC solve the same reduced QP~\eqref{eq:reduced_qp} and
differ only in the residual used to form the reduced gradient
$g_{x,k} = J_k^\top e_k$: the model residual $M(u)-T^*$ for MO
and the plant residual $P(u)-T^*$ for ILC. The LF model runs
cooler and loses heat faster than the plant
($k_\text{LF}=15$ vs.\ $k_\text{HF}=8$~W/(m$\cdot$K), no latent
heat), so the two residuals lie on opposite sides of $T^*$ and
the resulting gradients point in opposite directions. MO raises
the energy input (higher powers, slower scan, shorter lag) to
correct an error present only in its model; on the plant the
excess energy holds the cell hot, producing the slow-cooling
trace in Figure~\ref{fig:profiles}. ILC, with the residual
evaluated on the plant, lowers the energy input instead
(Figure~\ref{fig:perseed}). MO converges to a minimizer of the
model cost, and the model error appears unattenuated in the
plant response at that solution. In ILC the model only
determines the search direction; because the residual is
evaluated on the plant, errors in the predicted thermal response
are compensated from one iteration to the next, and the final
iterate reflects plant measurements rather than model
predictions. The method therefore tolerates substantial model
error, provided the model gradients remain descent directions
for the plant cost.

\section{Non-single-line geometry}
\label{sec:results:zigzag}

\begin{figure}[ht]
    \centering
    \includegraphics[width=0.95\columnwidth]{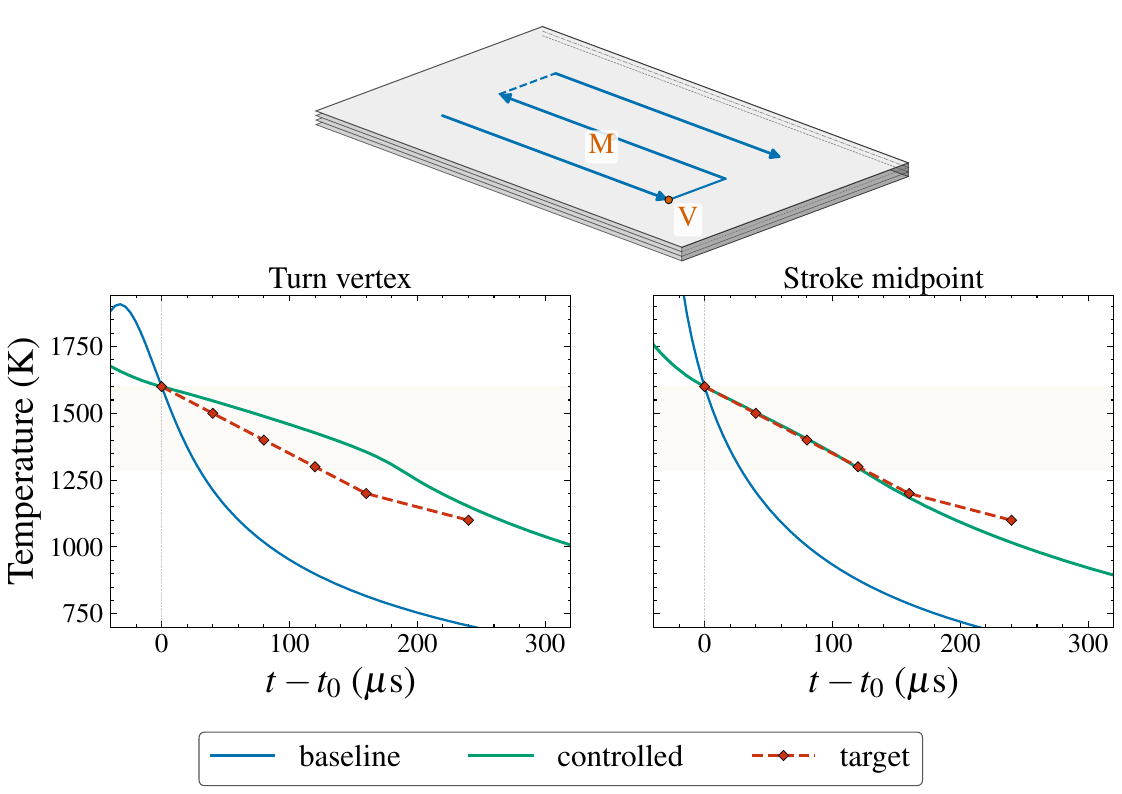}
    \caption{Transfer of ILC-optimized parameters to a three-stroke
zigzag scan with a velocity profile through corners. Top: geometry
with two monitors -- stroke midpoint (M) and turn vertex (V).
Bottom: $t_0$-aligned thermal response at each monitor for baseline
(blue) and ILC-controlled (green); target control points are red
diamonds along the dashed target line, solidification region
(1290--1600~K) shaded. Both monitors track the target cooling
profile under control; the baseline cools too fast. Peaks
(off-screen, left of $t_0$) are higher under control because the
trailing laser adds energy to an already-hot leading-laser pass;
cooling rate, not peak temperature, is the controlled quantity.}
    \label{fig:zigzag}
\end{figure}

AM scan paths are not globally repetitive: ILC's iteration-domain
formulation assumes the same scan is repeated each trial, while in
practice each layer is composed of strokes joined by corners and
direction reversals. We use a three-stroke zigzag
(Figure~\ref{fig:zigzag}) with layer stacking as a robustness check:
we hold the single-line ILC parameters fixed and apply them to the
zigzag without re-optimizing, and we monitor the thermal response at
two locations along the path: a stroke midpoint (M, representative
of the straight section) and a turn vertex (V, the corner that
accumulates the most heat from the scan-direction reversal) -- to
test whether the qualitative improvement still holds.

At both monitors the controlled scan tracks the target cooling
profile through the solidification region more closely than the
single-laser baseline, which cools too fast. The controlled scan
extends the time spent in the solidification region by 480\% at
the stroke midpoint and 364\% at the turn vertex relative to the
baseline. The vertex is the worst case for transfer where heat
accumulates at the scan-direction reversal, and the controlled
response runs hotter than the target there
(Figure~\ref{fig:zigzag}), but remains far closer to the
target profile than the baseline. Peak temperatures under
control are higher than the baseline at both locations but peak temperature is not the controlled quantity, and the
improvement appears in the cooling rate through the
solidification region, which is. The parameters were optimized
on a single straight line and applied to the zigzag without
re-optimization, showing the controller's benefit is not
specific to the straight-line setup it was tuned on.

\section{Conclusion}
\label{sec:conclusion}

We presented an SQP-based iterative learning control framework
for dual-laser LPBF that tracks a cooling profile favouring equiaxed grains
by optimizing the leading and trailing laser powers, lag
distance, and scan velocity. Across \nseeds{} initial conditions,
ILC reaches a tight band of plant solutions (CV \ilccv{}\%)
within roughly 20 iterations, despite realistic mismatch in
absorption, latent heat treatment, and powder-bed conductivity.
Model-only optimization converges to the LF optimum with parameter
selections that overshoot the target cooling profile when
evaluated on the plant. The framework requires no algorithmic
changes to deploy on hardware with pyrometer feedback. Future
work includes experimental validation and extension to spatially
varying control along the scan path.

\bibliographystyle{ieeetr}
\bibliography{bib}

@article{debroy2018additive,
  title={Additive manufacturing of metallic components--process, structure and properties},
  author={DebRoy, Tarasankar and Wei, Huiliang L and Zuback, James S and Mukherjee, Tuhin and Elmer, John W and Milewski, John O and Beese, Allison Michelle and Wilson-Heid, Alexander and De, Amitava and Zhang, Wei},
  journal={Progress in materials science},
  volume={92},
  pages={112--224},
  year={2018},
  publisher={Elsevier}
}

@article{chia2022process,
  title={Process parameter optimization of metal additive manufacturing: A review and outlook},
  author={Chia, Hou Yi and Wu, Jianzhao and Wang, Xinzhi and Yan, Wentao},
  journal={Journal of Materials Informatics},
  volume={2},
  number={4},
  pages={N--A},
  year={2022},
  publisher={OAE Publishing Inc.}
}

@article{kurz2001columnar,
  title={Columnar to equiaxed transition in solidification processing},
  author={Kurz, Wilfried and Bezen{\c{c}}on, C and G{\"a}umann, M},
  journal={Science and technology of advanced materials},
  volume={2},
  number={1},
  pages={185--191},
  year={2001}
}

@article{hunt1984steady,
  title={Steady state columnar and equiaxed growth of dendrites and eutectic},
  author={Hunt, James D},
  journal={Materials science and engineering},
  volume={65},
  number={1},
  pages={75--83},
  year={1984},
  publisher={Elsevier}
}

@article{gaumann2001single,
  title={Single-crystal laser deposition of superalloys: processing--microstructure maps},
  author={G{\"a}umann, Matthias and Bezen{\c{c}}on, Cyrille and Canalis, P and Kurz, Wilfried},
  journal={Acta materialia},
  volume={49},
  number={6},
  pages={1051--1062},
  year={2001},
  publisher={Elsevier}
}

@article{zhang2022process,
  title={Process parameters optimisation for mitigating residual stress in dual-laser beam powder bed fusion additive manufacturing},
  author={Zhang, Wenyou and Abbott, William M and Sasnauskas, Arnoldas and Lupoi, Rocco},
  journal={Metals},
  volume={12},
  number={3},
  pages={420},
  year={2022},
  publisher={MDPI}
}

@article{bergmueller2023enhancing,
  title={Enhancing equiaxed grain formation in a high-alloy tool steel using dual laser powder bed fusion},
  author={Bergmueller, Simon and Scheiber, Josef and Kaserer, Lukas and Leichtfried, Gerhard},
  journal={Additive Manufacturing},
  volume={74},
  pages={103727},
  year={2023},
  publisher={Elsevier}
}

@article{vanini2025local,
  title={Local microstructure engineering of super duplex stainless steel via dual laser powder bed fusion--an analytical modeling and experimental approach},
  author={Vanini, Michele and Searle, Samuel and Vanmunster, Lars and Vanmeensel, Kim and Vrancken, Bey},
  journal={Additive Manufacturing},
  pages={104994},
  year={2025},
  publisher={Elsevier}
}

@article{shahriari2015taking,
  title={Taking the human out of the loop: A review of Bayesian optimization},
  author={Shahriari, Bobak and Swersky, Kevin and Wang, Ziyu and Adams, Ryan P and De Freitas, Nando},
  journal={Proceedings of the IEEE},
  volume={104},
  number={1},
  pages={148--175},
  year={2015},
  publisher={IEEE}
}

@article{xue2023jax,
  title={JAX-FEM: A differentiable GPU-accelerated 3D finite element solver for automatic inverse design and mechanistic data science},
  author={Xue, Tianju and Liao, Shuheng and Gan, Zhengtao and Park, Chanwook and Xie, Xiaoyu and Liu, Wing Kam and Cao, Jian},
  journal={Computer Physics Communications},
  volume={291},
  pages={108802},
  year={2023},
  publisher={Elsevier}
}

@article{bristow2006survey,
  title={A survey of iterative learning control},
  author={Bristow, Douglas A and Tharayil, Marina and Alleyne, Andrew G},
  journal={IEEE control systems magazine},
  volume={26},
  number={3},
  pages={96--114},
  year={2006},
  publisher={Ieee}
}

@article{wang2009survey,
  title={Survey on iterative learning control, repetitive control, and run-to-run control},
  author={Wang, Youqing and Gao, Furong and Doyle III, Francis J},
  journal={Journal of process control},
  volume={19},
  number={10},
  pages={1589--1600},
  year={2009},
  publisher={Elsevier}
}

@article{owens2005iterative,
  title={Iterative learning control—An optimization paradigm},
  author={Owens, David H and H{\"a}t{\"o}nen, Jari},
  journal={Annual reviews in control},
  volume={29},
  number={1},
  pages={57--70},
  year={2005},
  publisher={Elsevier}
}

@article{mishra2010optimization,
  title={Optimization-based constrained iterative learning control},
  author={Mishra, Sandipan and Topcu, Ufuk and Tomizuka, Masayoshi},
  journal={IEEE Transactions on Control Systems Technology},
  volume={19},
  number={6},
  pages={1613--1621},
  year={2010},
  publisher={IEEE}
}

@article{liao2022robustness,
  title={On robustness in optimization-based constrained iterative learning control},
  author={Liao-McPherson, Dominic and Balta, Efe C and Rupenyan, Alisa and Lygeros, John},
  journal={IEEE Control Systems Letters},
  volume={6},
  pages={2846--2851},
  year={2022},
  publisher={IEEE}
}

@inproceedings{spector2018passivity,
  title={Passivity-based iterative learning control design for selective laser melting},
  author={Spector, Michael JB and Guo, Yijie and Roy, Souvik and Bloomfield, Max O and Maniatty, Antoinette and Mishra, Sandipan},
  booktitle={2018 Annual American Control Conference (ACC)},
  pages={5618--5625},
  year={2018},
  organization={IEEE}
}

@inproceedings{inyang2022model,
  title={Model-free multi-objective iterative learning control for selective laser melting},
  author={Inyang-Udoh, Uduak and Hu, Ruixiong and Mishra, Sandipan and Wen, John and Maniatty, Antoinette},
  booktitle={2022 American Control Conference (ACC)},
  pages={2879--2885},
  year={2022},
  organization={IEEE}
}

@article{liao2024layer,
  title={Layer-to-layer melt pool control in laser powder bed fusion},
  author={Liao-McPherson, Dominic and Balta, Efe C and Afrasiabi, Mohamadreza and Rupenyan, Alisa and Bambach, Markus and Lygeros, John},
  journal={IEEE transactions on control systems technology},
  volume={33},
  number={1},
  pages={207--218},
  year={2024},
  publisher={IEEE}
}

@article{kavas2025layer,
  title={Layer-to-layer Closed-loop Switched Heating and Cooling Control of the Laser Powder Bed Fusion Process},
  author={Kavas, Bar{\i}{\c{s}} and Balta, Efe C and Witte, Lars and Tucker, Michael R and Lygeros, John and Bambach, Markus},
  journal={arXiv preprint arXiv:2512.17518},
  year={2025}
}

@article{balta2024iterative,
  title={Iterative learning spatial height control for layerwise processes},
  author={Balta, Efe C and Tilbury, Dawn M and Barton, Kira},
  journal={Automatica},
  volume={167},
  pages={111756},
  year={2024},
  publisher={Elsevier}
}

@article{trapp2017situ,
  title={In situ absorptivity measurements of metallic powders during laser powder-bed fusion additive manufacturing},
  author={Trapp, Johannes and Rubenchik, Alexander M and Guss, Gabe and Matthews, Manyalibo J},
  journal={Applied Materials Today},
  volume={9},
  pages={341--349},
  year={2017},
  publisher={Elsevier}
}

@article{zhang2019thermal,
  title={On thermal properties of metallic powder in laser powder bed fusion additive manufacturing},
  author={Zhang, Shanshan and Lane, Brandon and Whiting, Justin and Chou, Kevin},
  journal={Journal of manufacturing processes},
  volume={47},
  pages={382--392},
  year={2019},
  publisher={Elsevier}
}

@article{marchese2024heat,
  title={Heat-treated Inconel 625 by laser powder bed fusion: microstructure, tensile properties, and residual stress evolution},
  author={Marchese, Giulio and Piscopo, Gabriele and Lerda, Serena and Salmi, Alessandro and Atzeni, Eleonora and Biamino, Sara},
  journal={Journal of Materials Engineering and Performance},
  volume={33},
  number={13},
  pages={6825--6834},
  year={2024},
  publisher={Springer}
}

@inproceedings{balula2023sequential,
  title={Sequential quadratic programming-based iterative learning control for nonlinear systems},
  author={Balula, Samuel and Balta, Efe C and Liao-McPherson, Dominic and Rupenyan, Alisa and Lygeros, John},
  booktitle={2023 IEEE Conference on Control Technology and Applications (CCTA)},
  pages={162--167},
  year={2023},
  organization={IEEE}
}

\end{document}